# Onboard Dynamic Rail Track Safety Monitoring System

[1]Abhisekh Jain S [2]Arvind S [3]Balaji B.S [4]Ram Viyas N.P
(1) III YEAR B.E. EEE, (2) & (3) III YEAR B.E. ECE, (4) III YEAR B.E. EIE
Thiagarajar College of Engineering, Madurai, India.

## *ABSTRACT*

*This proposal aims at solving one of the long prevailing problems in the Railways. This simple method of continuous monitoring and assessment of the condition of the rail tracks can prevent major disasters and save precious human lives. Our method is capable of alerting the train in case of any dislocations in the track or change in strength of the soil. Also it can avert the collisions of the train with other or with the vehicles trying to move across the unmanned level crossings.*

## I. INTRODUCTION

Travel is fascinating, of them train travel is more exciting. With the increased comfort levels in train transport and traffic in trains we are in an extremely important situation to improve the safety concerns in train travel. This paper explores the possible ways of increasing train track safety using onboard monitoring and dynamic track monitoring system. History shows that many train accidents have occurred because of track failure the recent one being in the year 2005 near Hyderabad where seven bogies of an express derailed over a bridge and many lives were lost.

India with its increased technological inputs is in a demanding situation to counter this problem. Emulating the previous human based erroneous and procedural system where track monitoring is done twice in a year, our dynamic technology will enable the trains to run safely. The previous system had only track monitoring and no onboard transformation of the data dynamically, but this paper proves its worth by providing solutions for the same. This paper also explores the possible future in track monitoring and implementation of the system in a global scale. These systems are cost worth systems and on implementation can yield excellent results.

India leading in the train track distance all over the world and with its varsity will be the best place to implement the system. Note worthy things are dynamic speed control and user friendliness of this system. The system tackles almost all possible failure modes including collisions and how to counter them by the application of emergency brakes in the train. With the use of high end RF system the consistency of the system is not a major concern.

This system comes in two challenging models one ofhe version gets its power from solar and another from power line. Implementation is based on the viability of the area in which t the sensor is deployed. Data losses are almost negligible in this system. For high traffic tracks, faster train monitoring are enabled using separate vibratory circuits placed at suitable areas to increase the consistency of the system and correlate the results.

## II. BLOCK DIAGRAM

This alerting system mainly comprises of two modules, the first one is the sensor network that monitors the track before the train crosses and the other is the wireless network that receives the information from the sensor network and alerts the train well in advance about the impending disaster. This method can process the information from the sensor network and alert the train giving it enough time to stop and hence prevent the disaster.

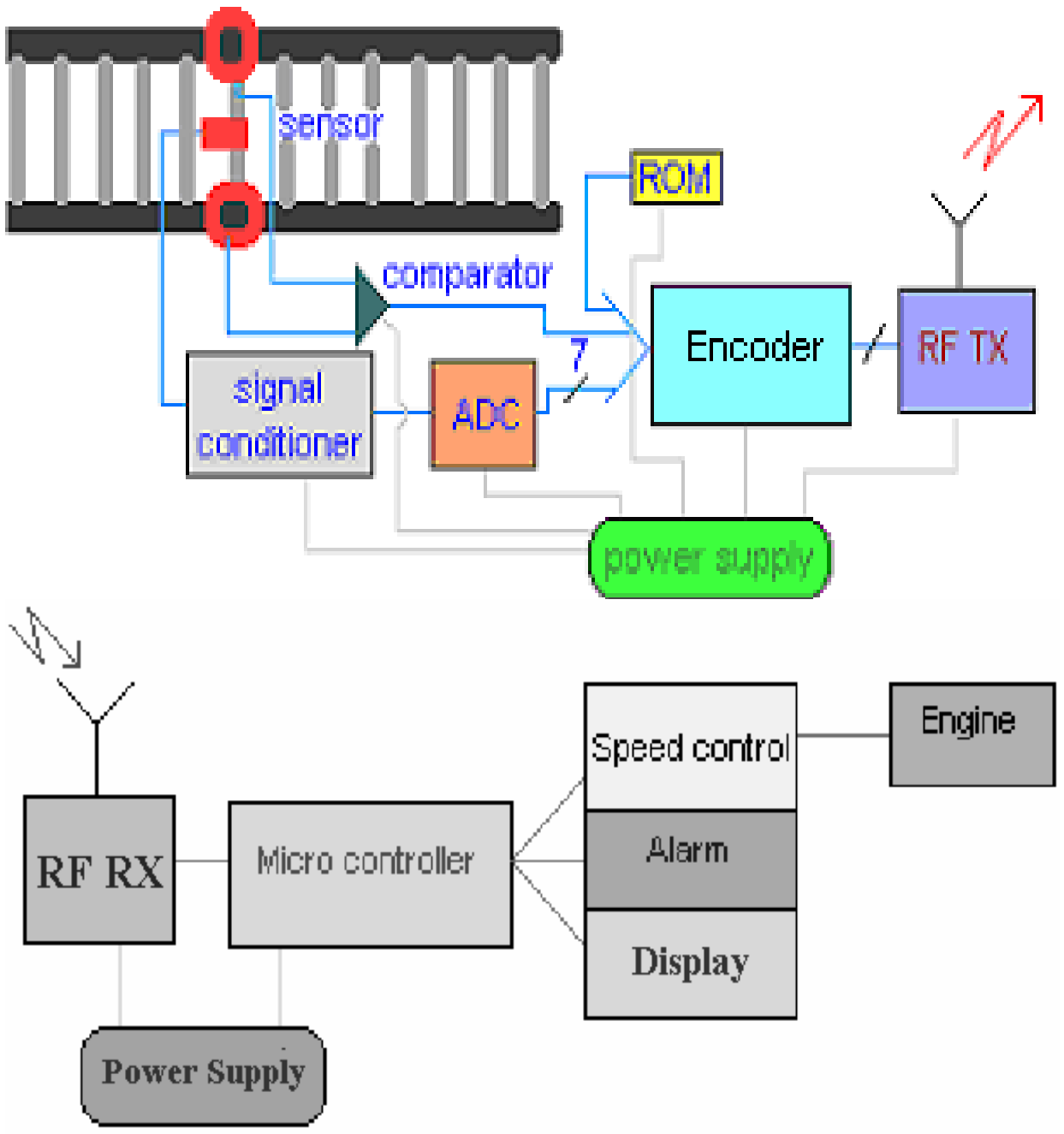

Fig. 1. Block Diagram

### Sensor Network

Our network involves placing a group of threesensors near the tracks to monitor it before train crosses it. This group of sensors will be placed at regular intervals which in turn dependon the sensitivity of sensors used. The group of sensors consists of two piezo-electric sensors which are placed on either side of the track and another sensor placed in the soil at a depth of 5 inches.

The Piezoelectric transducer used can be a lead-zirconate-titanate (PZT) piezoelectric pad. Along each node in the track a RF transmitter is placed to send the details collected by the Sensor Network. The transmitter used is SRT (Short Range Transmitter).

**Wireless Network**

This network consists of an RF transmitter placed near the Sensor Network and an RF transceiver in the train to collect the data from the transmitter. The received information is processed and checked for consistency with preloaded details. This process is done using a microcontroller. The resulting output is used to alert and control the speed mechanism of the train.

## III. WORKING

Piezoelectric is a low-cost material that experience strain when a voltage potential is applied across them; similarly, the material outputs a voltage when strained. The piezoelectric transducer used senses the vibration of the track and activates the sensor network. The sensors placed on either side of the tracks collects the existing details of the track, the outputs of which are fed to the Operational amplifier based comparator circuit. The resulting output (1bit) along with the information from the piezo-electric transducer (7bits) is given to an ADC along with address of the node. The digital output from the ADC is given to the encoder circuit which performs the Linear Block Coding. The RF receiver used can be Medium Range Receiver like AC4490, the range of which is 1.6 Km.

The details of vibrations of the track and of the nearby soil are sensed by these sensors and the data from those are digitized by an ADC and the net output is an Eight Bit Data. Now this information is to be transmitted so that the microcontroller in the train can process, thereby it can asses the grade of the track and it can adjust the speed of the train or can stop the train if there is any drastic change.

**Data Acquisition**

The vibrations in the track can be sensed before a distance that differs based on the profile speed which can be any where between (1-2Km), a level that can be sensed by our sensor. Once the Sensor senses the vibration for the first time the rest of the circuit is activated by signal from the vibrations received. For the sensor to get activated to this level the Train must be at least a particular distance say 1.6 Kilometers away from the sensor node. Once the circuit is activated immediately the data from the sensor is read and it is transmitted continually. By this time of just milliseconds the train would not have moved so much distance, thus for a train at that distance, the Microcontroller receiving this Data can process it.

The two sensors on the side of the tracks sense the vibrations in the two rails when the train approaches. Thus thereis not going to be large difference between the vibrations occurringin the two rails unless there is a Curve or there is some damage made to the track so that the vibrations differ by a very large amount. Thus the vibration levels occurring at the two rails are compared using a comparator after converting to suitable voltage level. This is considered as one bit message via, Goodness Flag, which is considered along with the other sensor's output.

The sensors used are piezoelectric sensors that are capable of sensing even vibrations of weak order. The sensitivity of the sensor adapted is chosen appropriately to fit this application. The comparator used is an ordinary single bit comparator using an Operational Amplifier since the power consumption is less. The data from the third sensor is converted into digital form by an Eight Bit ADC. The Seven Most Significant bits are taken and the eighth bit is the Goodness Flag (the output from the former mentioned sensor).

**Encoding Mechanism**

The transmission is going to be through RF Signals. Thus a suitable Wireless Transmitter is used. The Frequency of transmission can be chosen as any free ISM band or licensed band can also be used. The data is first encoded using a suitable Linear Block Code of appropriate weight. The Weights of the codes are chosen so that the error introduced in the transfer does not affect appreciably the results. Now the data is of eight bit length so that two bit accuracy is sufficient for deciding the speed to an accurate level thus a (29, 12) LBC is sufficient. The reason for using LBC is that they can be decoded easily thus providing the result fast enough that the train does not cross the sensor.

The reception must be done carefully since the train moves with very high velocity the Doppler Effect cannot be neglected so there is going to be a shift in frequency which has to be taken care. To do this we need to know the speed of the train which is got from the System present. Now this information is given to the receiver which can make appropriate decisions. Once the signal is received the Data is extracted from this signal then it is fed to the Microcontroller. It decodes the transmitted signal and the errors if any are corrected. Now the data from the sensors is available to the microcontroller.

Now the bit of the Goodness Flag has to be processed. The Microcontroller can check the status of the Track. Therefore only if the Track is safe the controller proceeds to the next seven bits of data for speed calculation purpose, else it waits to have a conformation of possible Track Failure. This is got from consistent Failure messages from the same or successive nodes. So the Microcontroller waits until next information comes from that sensor or from the next sensor node to make decision.

Once consistent information arrived conveys possible track failure, the Microcontroller sends alarm to the driver to stop the train. If consistent Information does not say that the track is damaged, the Microcontroller warns the Driver so that one can see the track visually and inspect it if needed. So the possibility of making wrong decisions is very low.

**Control Technique**

If there is no problem in the track i.e. if theGoodness

Flag is not set, then the microcontroller computes the needed speed the train should travel in that part of the track, and as it knows the present speed of the train, it gives how much acceleration or deceleration is to be done to the Engine control, thereby making the speed control. This signal is the error signal that is corrective to engine's speed. This makes a closed loop control system there by regulating the speed of the train by measuring the vibrations due to the train itself.

The Sensors are going to be spaced by an optimum distance so that enough information about that part of track is given to the Train. If there is some rapid fluctuations of the readings the engine controller can be implemented as a P.I.D. Controller so that the time rate of change is also taken into account and the speed can be adjusted to a maximum level of satisfaction.

## IV. ADDRESSING THE NODES

The Sensory nodes are going to be placed at regular distances in the derailment prone areas. Thus addressing of the nodes is done in the following way. On the journey of the train the first node is given the address zero, the second node one and is repeated for up to fifteen. Once fifteen is reached again we start with zero. Once a fifteen address is received a count is maintained in the train. Thus the train may also calculate its approximate speed from the time of arrival of the information from the next sensor node. If the data from any sensor node gets lost then it can be identified from the address of the next received data set. Thus we can even detect the malfunctioning of some of the sensors nodes if it has been damaged. Thus a four bit address is enough for a node. This address is stored on chip in a ROM and is encoded along with the data bits and is transmitted. The ROM is programmable so the address can be changed easily.

## V. SIGNAL CONDITIONING AND POWER REQUIREMENTS

The electrical signals from the sensor are crude so it needs to be conditioned so we use op–amp based signal conditioners. The amplitude should not go beyond a particular level; it must not load the source. These things are taken care by the Signal Conditioner.

The power to the entire circuit is provided from the stored energy of rechargeable Lithium ion Batteries. A voltage of five volts at 500mW is sufficient. The power can be tapped from the nearby Rail Traffic Signal Post. By this way we can also detect the arrival of train from the occurrence of green signal and also in the remote areas, a battery is charged by a Solar Panel. The Power Budget is designed carefully and the components are kept within the limits. The portability and feasibility of this is that the sensors must touch the rails they are available in the form of thin sheets so that it can be stuck. The wireless module can be made as a System on Chip.

## VI. STATISTICAL ANALYSIS

### Derailment Prone Areas (1985-2006)

The Highlighted areas are derailment prone areas according to the survey of accidents during the year 1985 to 2006 in India.

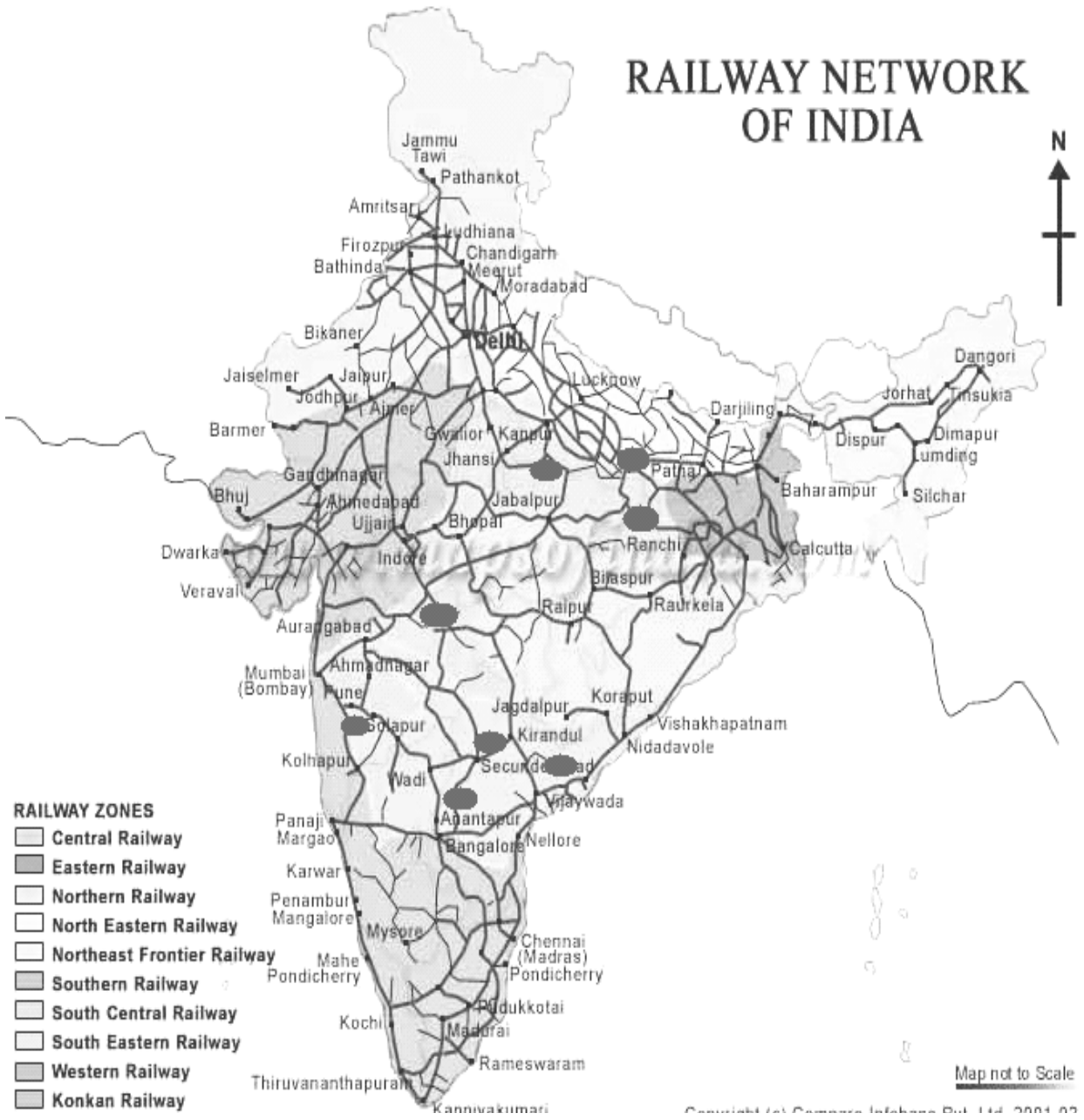

Fig. 2. Derailment Prone AreasCourtesy: www.hindustantimes.com

### Stastical Data of Major Train Accidents: An Eye Opener for Implementing the Project

TABLE I
STATISTICAL DATA OF MAJOR TRAIN ACCIDENTS

| Cause | Number of People Affected | Number of Accidents |
|---|---|---|
| Fire | 160 | 3 |
| Derailment | 1221 | 14 |
| Collision | 790 | 11 |
| Human errors | 386 | 4 |
| Others | 448 | 8 |
| Bomb blast | 226 | 3 |

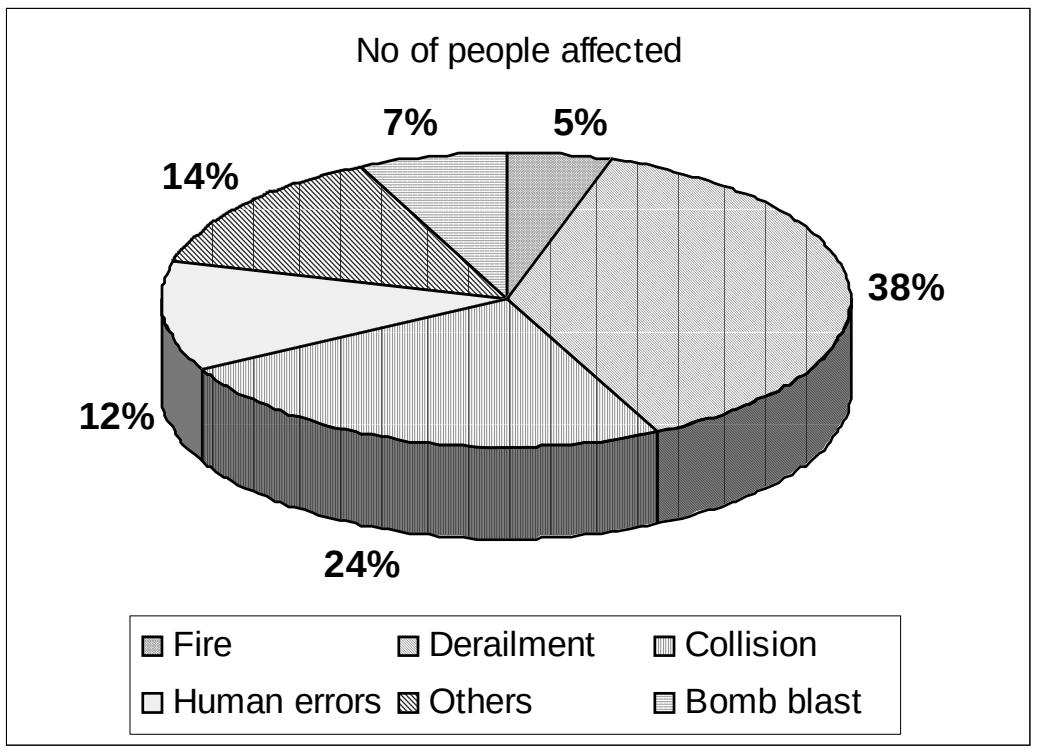

Fig. 3. Cause for the Accidents versus Number of People Affected

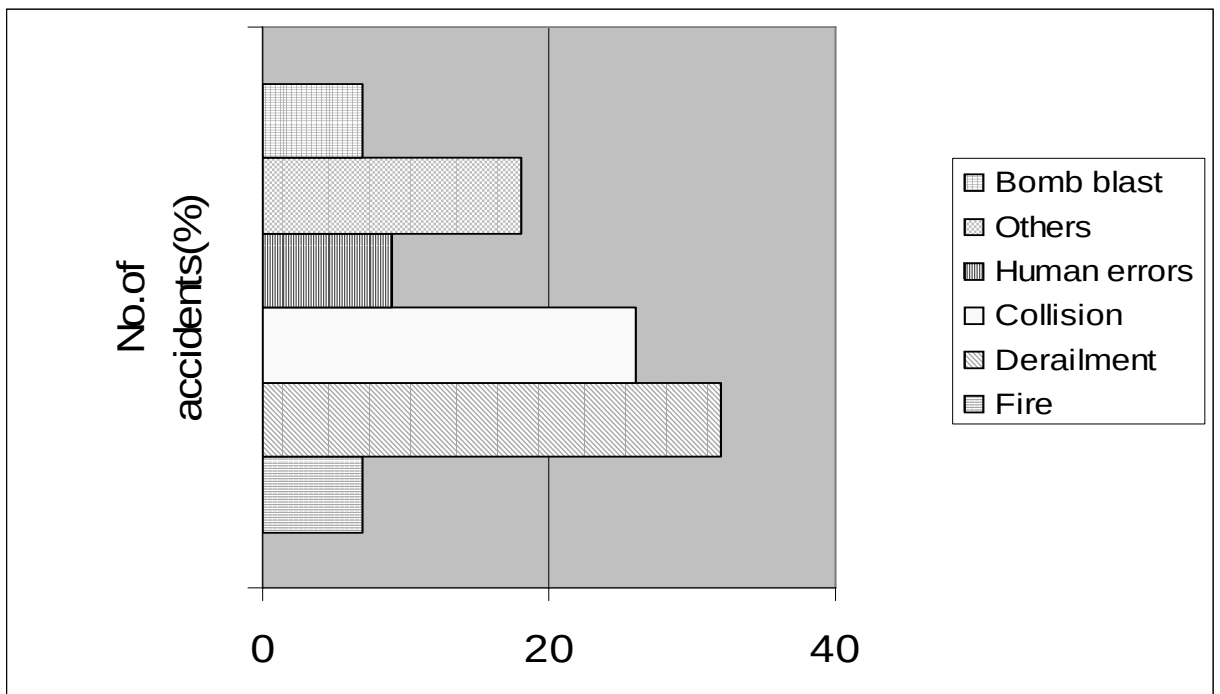

| | No.of accidents(%) |
|---|---|
| Bomb blast | 7 |
| Others | 18 |
| Human errors | 9 |
| Collision | 26 |
| Derailment | 32 |
| Fire | 7 |

Fig. 4. Causes for the Accidents versus Number of Accidents

## VII. CONCLUSION

Innovation is needed but more important is the application of existing technology. This treatise is one such attempt to use the wireless technology for providing safety to the people. Due to its dynamic nature the solution when applied will continue to yield good results paving way for innovations in the field of railway transport including levitated and bullet trains increasing train safety and importantly satisfied customers.

This work is dedicated to all the people who have lost their life in various train accidents.